\documentclass[aps,pra,10pt,showpacs,amsmath,twocolumn,floatfix]{revtex4-1}
\usepackage{amsmath}
\usepackage{amsfonts}
\usepackage{graphicx}

\newcommand{\diff}[2]{\frac{d #1}{d #2}}

\newcommand{\avg}[1]{\langle#1\rangle}
\newcommand{\Avg}[1]{\left\langle#1\right\rangle}

\newcommand{\bk}[1]{\left(#1\right)}
\newcommand{\Bk}[1]{\left[#1\right]}

\begin{document}
\title{Cavity quantum electro-optics}

\author{Mankei Tsang}

\email{mankei@unm.edu}

\affiliation{
Center for Quantum Information and Control,
University of New Mexico, MSC07--4220, Albuquerque, New Mexico
87131-0001, USA}








\date{\today}

\begin{abstract}
  The quantum dynamics of the coupling between a cavity optical field
  and a resonator microwave field via the electro-optic effect is
  studied. This coupling has the same form as the opto-mechanical
  coupling via radiation pressure, so all previously considered
  opto-mechanical effects can in principle be observed in
  electro-optic systems as well.  In particular, I point out the
  possibilities of laser cooling of the microwave mode, entanglement
  between the optical mode and the microwave mode via electro-optic
  parametric amplification, and back-action-evading optical
  measurements of a microwave quadrature.
\end{abstract}

\maketitle
\section{Introduction}
Recent technological advances in optics \cite{optics,optomechanics},
mechanics \cite{mechanics,optomechanics}, and superconducting
microwave circuits \cite{microwave} have made observations of various
quantum phenomena in microscopic systems possible. Such advances also
suggest that quantum effects will play a major role in future
communication, sensing, and computing technology. Each physical system
has its own advantages and disadvantages for different applications,
so it is desirable to transfer classical or quantum information to
different systems for different purposes. At the quantum level,
accurate information transfer requires coherent coupling among the
systems. While opto-mechanical and electro-mechanical coupling have
attracted a lot of attention recently \cite{mechanics,optomechanics},
the coherent coupling between optical and microwave fields has been
largely overlooked in the context of quantum information processing,
even though electro-optic modulation is a well known phenomenon in
classical optics \cite{yariv}. Previous studies on the quantum aspects
of the electro-optic effect mainly treat the microwave as a classical
signal for linear quantum optical processing \cite{electrooptics}, but
treating both fields quantum-mechanically should give rise to novel
physics.

In this paper, I consider both the optical field and the microwave
field as quantum degrees of freedom and study the coupling between the
fields via the electro-optic effect. Both fields are assumed to be
modes in a cavity or a resonator, since it is likely that both of them
need to be resonantly enhanced for quantum effects to be
observable. In terms of prior work, Ilchenko \textit{et al.}\ have
previously outlined such a theory in their study of photonic microwave
receivers \cite{ilchenko}, while Matsko \textit{et al.}\ have studied
the quantum back-action noise in electro-optic modulation
\cite{matsko}.  As noted by Matsko \textit{et al.}, the electro-optic
coupling has the same form as the opto-mechanical coupling via
radiation pressure, so all previously considered opto-mechanical
effect can in principle be observed in electro-optic systems as
well. In particular, I point out the possibilities of laser cooling of
the microwave mode, entanglement between the optical mode and the
microwave mode via electro-optic parametric amplification, and
back-action-evading optical measurements of a microwave
quadrature. All these effects require optical sideband pumping, which
is not considered in Refs.~\cite{ilchenko,matsko} and is the main
technical novelty of this paper compared with the prior work on
quantum electro-optics. If realized, the proposed effects should be
useful for both fundamental science and information processing
applications in the quantum regime. For example, electro-optic cooling
allows one to observe quantum optical effects at higher background
temperatures or lower microwave or even radio frequencies, while
coherent optical detection of microwave quadratures potentially allows
one to leverage the high efficiency of optical detectors compared with
conventional microwave detectors, the highest reported quantum
efficiency of which is only 27\% \cite{teufel}, for continuous
low-noise microwave measurements.

\section{\label{formalism}Formalism}

\begin{figure}[htbp]
\centerline{\includegraphics[width=0.5\textwidth]{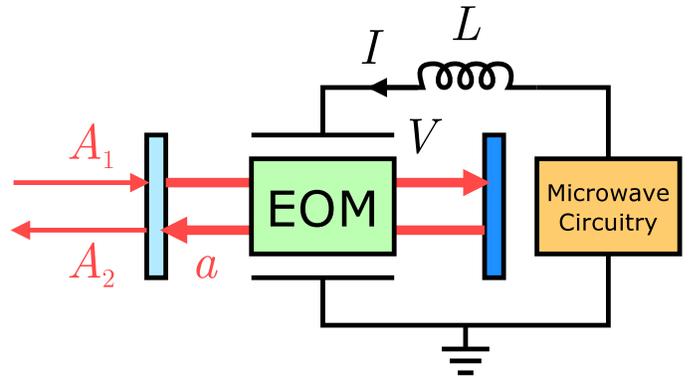}}
\caption{(Color online). A basic cavity electro-optic system. EOM
  denotes an electro-optic modulator.}
\label{oe_setup}
\end{figure}

A full quantum treatment of the electro-optic
effect can be done using standard quantum optics theory
\cite{yariv,mandel}, but here I shall use a more phenomenological
treatment to emphasize the similarity between electro-optics and
opto-mechanics. Consider the cavity electro-optic system depicted in
Fig.~\ref{oe_setup}. The transverse electro-optic modulator, which
consists of a second-order nonlinear optical medium, such as lithium
niobate or an electro-optic polymer, introduces a voltage-dependent
phase shift to the intra-cavity optical field resonant at frequency
$\omega_a$. For simplicity, I first consider one optical mode.  It is
straightforward to show that the interaction Hamiltonian is
\begin{align}
H_i &= -\frac{\hbar}{\tau} \phi a^\dagger a,
\end{align}
where $a$ and $a^\dagger$ are the optical annihilation and creation
operators, respectively, which obey the commutation relation
$[a,a^\dagger] = 1$, $\tau$ is the optical round-trip time, and $\phi$
is the single-round-trip phase shift. This Hamiltonian is exactly the
same as the one for opto-mechanics, in which case $\phi$ is the phase
shift introduced by the cavity mirror displacement
\cite{optomechanics}. The round-trip electro-optic phase shift, on the
other hand, is given by \cite{yariv}
\begin{align}
\phi &= \frac{\omega_a n^3 r l}{c d} V,
\end{align}
where $n$ is the optical refractive index inside the electro-optic
medium, $r$ is the electro-optic coefficient in units of m$/$V, $l$ is
the length of the medium along the optical axis, $d$ is the thickness,
and $V$ is the voltage across the medium. If the modulator is modeled
as a capacitor in a single-mode microwave resonator, one can quantize
the voltage by defining
\begin{align}
V &= \bk{\frac{\hbar\omega_b}{2C}}^{1/2}\bk{b + b^\dagger},
\end{align}
where $b$ and $b^\dagger$ are the microwave annihilation and creation
operators, respectively, which obey $[b,b^\dagger]=1$, $\omega_b$ is the
microwave resonant frequency, and $C$ is the capacitance of the
microwave resonator.  The full Hamiltonian becomes
\begin{align}
H &= \hbar\omega_a a^\dagger a + \hbar\omega_b b^\dagger b
-\hbar g\bk{b + b^\dagger} a^\dagger a,
\\
g &\equiv 
\frac{\omega_a n^3 rl}{c\tau d}\bk{\frac{\hbar\omega_b}{2C}}^{1/2}.
\label{g}
\end{align}
The fact that the Hamiltonian for electro-optics has the same form as
that for opto-mechanics allows one to apply known results for the
latter to the former.  $V$ then plays the role of the mechanical
position and $I$, the current in the microwave resonator, plays the
role of the mechanical momentum.

For the whispering-gallery-mode electro-optic modulator reported by
Ilchenko \textit{et al.}\ \cite{ilchenko}, the microwave frequency
$\omega_b$ is close to the optical free spectral range $\Delta\omega$,
so one should include multiple optical modes in the analysis
\cite{ilchenko,matsko}. Dobrindt and Kippenberg also performed a
similar multi-mode analysis for opto-mechanics \cite{dobrindt}.  One
can limit the analysis to three optical modes if
$|\Delta\omega-\omega_b|$ is much larger than the optical linewidth
$\gamma_a$, so that higher-order optical harmonics are not resonantly
coupled, or if one introduces two or three non-degenerate optical
modes via normal-mode splitting using two or three coupled optical
cavities, so that higher-order modes are farther away in frequency
\cite{dobrindt,grudinin}.  To demonstrate cooling or parametric
amplification alone, only two modes are needed. One can also detune
the high-order modes by introducing an appropriate frequency
dispersion to the cavity. This may be done by engineering the geometry
of the cavity or using a photonic crystal structure.

Let $a$, $a_1$, and $a_2$ be the annihilation operators for the center
optical mode, red-detuned mode, and the blue-detuned mode,
respectively. The Hamiltonian becomes \cite{ilchenko,matsko}
\begin{align}
H &= \hbar\omega_a a^\dagger a + \hbar(\omega_a-\Delta\omega)a_1^\dagger a_1
+\hbar(\omega_a+\Delta\omega)a_2^\dagger a_2
\nonumber\\&\quad
+\hbar\omega_b b^\dagger b - \hbar g (b + b^\dagger)
(a + a_1 + a_2)^\dagger(a+a_1+a_2),
\label{multimodeH}
\end{align}
assuming that the coupling coefficients for different optical modes
are the same for simplicity and $g$ is slightly modified to account
for the partial overlap among the modes. In the following, I shall
focus on the experimentally more relevant case of multiple optical
modes.

\section{Cooling and parametric amplification}
\begin{figure}[htbp]
\centerline{\includegraphics[width=0.5\textwidth]{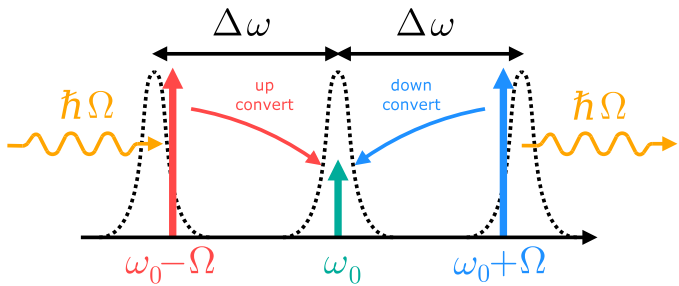}}
\caption{(Color online). Red-detuned optical pumping cools the
  microwave mode by transferring energy from the microwave to the
  optical mode at $\omega_a$ via parametric
  up-conversion. Blue-detuned pumping causes non-degenerate parametric
  down-conversion to the optical mode at $\omega_a$ and the microwave
  mode.}
\label{cool_pa}
\end{figure}

To describe cooling and parametric amplification, it is more
convenient to use the interaction picture. Let
\begin{align}
b &= \tilde{b}\exp(-i\omega_b t),
\\
a &= \tilde{a}\exp(-i\omega_a t),
\\
a_1 &= a_-\exp[-i(\omega_a - \Delta\omega)t],
\\
a_2 &= a_+\exp[-i(\omega_a+\Delta\omega)t].
\end{align}
With the rotating-wave approximation, the Hamiltonian becomes
\begin{align}
H_I &\approx H_C + H_A,
\\
H_C &\equiv -\hbar g \bk{\alpha_- \tilde{a}^\dagger \tilde{b} 
+\alpha_-^\dagger \tilde{a} \tilde{b}^\dagger},
\\
H_A &\equiv -\hbar g \bk{\alpha_+ \tilde{a}^\dagger \tilde{b}^\dagger
+ \alpha_+^\dagger \tilde{a} \tilde{b}},
\\
\alpha_- &\equiv a_- \exp[i(\Delta\omega-\omega_b)t]
= a_1 \exp[i(\omega_a-\omega_b)t],
\\
\alpha_+ &\equiv a_+ \exp[-i(\Delta\omega-\omega_b)t]
= a_2 \exp[i(\omega_a+\omega_b)t].
\end{align}
If one pumps the optical cavity at frequencies $\omega_a \pm\omega_b$
such that $\alpha_\pm$ can be approximated as classical complex
numbers, $H_C$ corresponds to linear coupling between $\tilde{a}$ and
$\tilde{b}$ like a beam splitter, while $H_A$ corresponds to
non-degenerate parametric amplification of the two modes. These two
effects are schematically described in Fig.~\ref{cool_pa}.

With red-detuned optical pumping at $\omega_a-\omega_b$, the microwave
field is linearly coupled to an extra optical reservoir, so the
microwave mode can be cooled. To account for coupling to travelling
waves and thermal reservoirs, the easiest way is to use quantum
Langevin equations \cite{mandel,zoller}.  Assuming $H_A \approx 0$,
the equations of motion for $\tilde{a}$ and $\tilde{b}$ are
\begin{align}
\diff{\tilde{a}}{t} &= 
ig \alpha_- \tilde{b}  -\frac{\gamma_a}{2} \tilde{a}+\sqrt{\gamma_a} A,
\\
\diff{\tilde{b}}{t} &= ig \alpha_-^* \tilde{a}- \frac{\gamma_b}{2} \tilde{b}  + 
 \sqrt{\gamma_b} B,
\end{align}
where $\gamma_a$ and $\gamma_b$ are the optical and microwave decay
coefficients, respectively, and $A$ and $B$ are input quantum Langevin
noise operators, which obey the commutation relations given by
\begin{align}
\Bk{A(t),A^\dagger(t')} &= \delta(t-t'),
&
\Bk{B(t),B^\dagger(t')} &= \delta(t-t').
\end{align}
Assuming thermal and white statistics for $A$ and $B$
such that
\begin{align}
\Avg{A^\dagger(t')A(t)} &= N(\omega_a)\delta(t-t'),
\\
\Avg{B^\dagger(t')B(t)} &= N(\omega_b)\delta(t-t'),
\\
N(\omega) &= \frac{1}{\exp(\hbar\omega/k_BT)-1},
\end{align}
it can be shown that the microwave occupation number at steady state
is
\begin{align}
\avg{\tilde{b}^\dagger \tilde{b}}_{t\to\infty} 
= \frac{N(\omega_b)+GN(\omega_a)}{1+G},
\\
G \equiv 
\frac{G_0}{1+(\gamma_b/\gamma_a)(1+
G_0)},
\quad
G_0 \equiv \frac{4g^2|\alpha_-|^2}{\gamma_a\gamma_b},
\end{align}
which implies cooling for $g \neq 0$, since $N(\omega_a) <
N(\omega_b)$ and $\avg{\tilde{b}^\dagger \tilde{b}}_{t\to\infty} <
N(\omega_b)$.  This is the electro-optic analog of opto-mechanical
sideband cooling \cite{cooling}.  Note that $G$ can saturate with
increasing pump power when $G_0 \sim \gamma_a/\gamma_b$ and the upper
limit is
\begin{align}
\lim_{G_0\to\infty}G &= \frac{\gamma_a}{\gamma_b},
\label{G_limit}
\end{align}
in which case cooling is limited by the decay rate of the optical mode
rather than the electro-optic coupling strength \cite{grajcar}. The
linear coupling between $\tilde a$ and $\tilde b$ also allows
classical or quantum information to be transferred coherently between
the microwave mode and the optical mode and may be useful as a
coherent microwave receiver \cite{ilchenko}.

With blue-detuned pumping at $\omega_a+\omega_b$, the parametric
amplification process can create entangled photons or, equivalently, a
two-mode squeezed state in the optical and microwave modes, if thermal
fluctuations are negligible.  The effect of thermal fluctuations on
the entanglement can also be studied using quantum Langevin equations
and is qualitatively the same as that on the analogous phenomenon of
opto-mechanical entanglement \cite{vitali}. One may also use
parametric amplification beyond threshold to generate coherent
microwaves, analogous to a phonon laser in opto-mechanics
\cite{rokhsari,grudinin}. It can be shown that the oscillation
threshold condition is $4g^2|\alpha_+|^2/(\gamma_a\gamma_b) \ge 1$,
which requires a similar number of pump photons to that required for
significant cooling.

\section{Effect of parasitic down-conversion on cooling}
\begin{figure}[htbp]
\centerline{\includegraphics[width=0.5\textwidth]{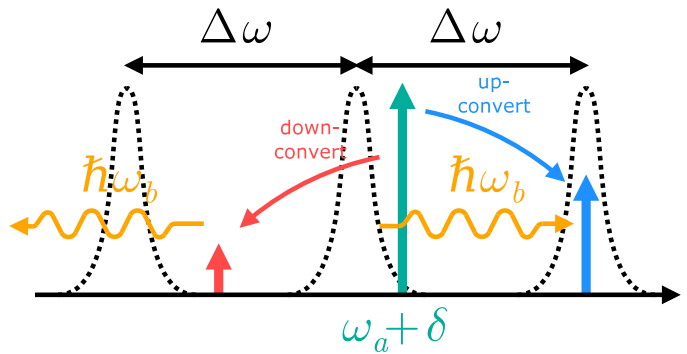}}
\caption{(Color online). Electro-optic cooling via detuned pumping of
  the center optical mode. The parasitic down-conversion process can
  limit the lowest microwave occupation number one can achieve.}
\label{higher-order}
\end{figure}

As mentioned in Sec.~\ref{formalism}, if a whispering-gallery-mode
optical cavity is used, parasitic coupling to higher-order modes can
hamper the cooling or parametric amplification efficiency. To
investigate the effect of this parasitic coupling on cooling, consider
optical pumping at frequency $\omega_a +\delta$ and $\delta =
\Delta\omega - \omega_b$, as shown in Fig.~\ref{higher-order}. The
up-conversion process is then resonantly enhanced with respect to the
down-conversion process.  Assuming undepleted pumping, the
steady-state center-mode field is
\begin{align}
\tilde a &\approx 
\alpha_0 \exp(-i\delta t),
&
\alpha_0 &\equiv \frac{\sqrt\gamma_a\avg{A}}
{-i\delta+\gamma_a/2}.
\end{align}
Defining
\begin{align}
a_- = \tilde a_-\exp(-2i\delta t),
\end{align}
the coupled-mode equations become
\begin{align}
\diff{\tilde a_-}{t} &= 2i\delta \tilde a_- + ig\alpha_0\tilde b^\dagger
-\frac{\gamma_a}{2} \tilde a_- + \sqrt{\gamma_a} A_-,
\label{am}\\
\diff{a_+}{t} &= ig\alpha_0\tilde b-\frac{\gamma_a}{2} a_++\sqrt{\gamma_a} A_+,
\label{ap}\\
\diff{\tilde b}{t} &= ig\alpha_0^* a_+ + ig\alpha_0 \tilde a_-^\dagger
-\frac{\gamma_b}{2}\tilde b + \sqrt{\gamma_b} B,
\label{b}
\end{align}
where $A_-$ and $A_+$ are the input Langevin noise operators for the
$\tilde a_-$ and $a_+$ modes, respectively. Assuming
\begin{align}
N(\omega_a\pm\Delta\omega) &\approx 0, &
\gamma_b&\ll \gamma_a,
&
2g|\alpha_0| &\ll \gamma_a
\end{align}
for simplicity, I obtain, after some
straightforward but cumbersome algebra,
\begin{align}
\avg{\tilde b^\dagger \tilde b}_{t\to\infty}
\approx \frac{N(\omega_b)+\Gamma\mu}
{1+\Gamma},
\label{nb}
\\
\Gamma \equiv
\frac{\Gamma_0}{1+\mu},
\quad
\Gamma_0 \equiv \frac{4g^2|\alpha_0|^2}{\gamma_a\gamma_b},
\quad
\mu \equiv \frac{\gamma_a^2}{16\delta^2}.
\end{align}
The lowest achievable microwave occupation number is thus
\begin{align}
\lim_{\Gamma_0\to\infty}\avg{\tilde b^\dagger \tilde b}_{t\to\infty} &=
\mu \equiv\frac{\gamma_a^2}{16\delta^2}.
\end{align}
Interestingly, this expression is identical to that for the minimum
mechanical occupation number in opto-mechanical cooling using one
optical mode \cite{cooling}. Since the pump photon number
$|\alpha_0|^2$ also depends on $\delta$, the optimal $\delta$ depends
on the specific parameters of the system and can be calculated by
minimizing $\avg{\tilde b^\dagger \tilde b}_{t\to\infty}$ with respect
to $\delta$.  For example, when $N(\omega_b) \gg \Gamma\mu$, $\Gamma$
is maximized and $\avg{\tilde b^\dagger \tilde b}_{t\to\infty}$ is
minimized when $\mu = 0.5$.

\section{Optical quantum non-demolition (QND) measurements of a
  microwave quadrature}
To perform QND measurements of the microwave voltage, one can pump the
optical cavity with laser light at $\omega_a$ in the
single-optical-mode case and estimate the microwave voltage by
measuring the phase of the output optical wave $A'=\sqrt{\gamma_a}
\tilde{a} - A$.  If the optical pump is strong enough, in addition to
the intrinsic shot noise in the phase quadrature of $A'$, one can also
observe excess noise due to the back-action of light on the
microwave. The shot noise relative to the signal decreases with
optical power but the back-action noise increases with power, so there
is a standard quantum limit (SQL) to voltage measurement sensitivity,
much like the SQL of displacement measurement due to radiation
pressure \cite{braginsky,matsko}. In both cases, the SQL can be
overcome using squeezed light \cite{caves,kimble,lowry}, variational
measurement \cite{kimble,vyatchanin}, or two mechanical or microwave
resonators \cite{briant,caniard}. Although the SQL has not been
observed in opto-mechanical or electro-optic systems, it can be
simulated by adding classical excess noise to the optical pump
\cite{lowry,caniard}, and the aforementioned noise cancellation
techniques can be used to remove any classical back-action noise as
well.

An interesting way of making back-action-evading measurements is to
use double-sideband optical pumping \cite{braginsky,hertzberg}.
Assuming equal-magnitude and undepleted double-sideband pumping, so
that
\begin{align}
\alpha_+ &= |\alpha|\exp(i\theta_+),
&
\alpha_- &= |\alpha|\exp(i\theta_-),
\end{align}
defining quadrature operators as
\begin{align}
X_a &\equiv \exp(-i\theta)\tilde{a} + \exp(i\theta) \tilde{a}^\dagger,
\\
Y_a &\equiv  -i\Bk{\exp(-i\theta)\tilde{a} - \exp(i\theta) \tilde{a}^\dagger},
\\
X_b &\equiv \exp(-i\nu)\tilde{b}+\exp(i\nu)\tilde{b}^\dagger,
\\
Y_b &\equiv -i\Bk{\exp(-i\nu)\tilde{b}-\exp(i\nu)\tilde{b}^\dagger},
\\
\xi &\equiv  \exp(-i\theta)A + \exp(i\theta) A^\dagger,
\\
\eta &\equiv -i\Bk{\exp(-i\theta)A - \exp(i\theta) A^\dagger},
\\
\theta &\equiv \frac{\theta_++\theta_-}{2},
\quad
\nu \equiv \frac{\theta_+-\theta_-}{2},
\end{align}
and again making the rotating-wave approximation, the equations
of motion become
\begin{align}
\diff{X_a}{t} &= -\frac{\gamma_a}{2} X_a + \sqrt{\gamma_a}\xi,
&
\diff{X_b}{t} &= 0,
\\
\diff{Y_a}{t} &= 2g|\alpha| X_b -\frac{\gamma_a}{2} Y_a + \sqrt{\gamma_a}\eta,
&
\diff{Y_b}{t} &= 2g|\alpha| X_a,
\end{align}
where coupling to travelling waves and reservoirs for the microwave
mode is neglected for clarity.  The measured microwave quadrature
$X_b$ is dynamically decoupled from the orthogonal quadrature $Y_b$,
so the back-action introduced to $Y_b$ via $X_a$ does not affect the
estimation accuracy of $X_b$. $X_b$ and $Y_a$ can be changed
independently by adjusting the phases of the two pump waves. This
technique should also enable microwave squeezing by measurement and
feedback \cite{wiseman}.  The double-sideband-pumping
back-action-evading measurement scheme has recently been demonstrated
for the measurement of a mechanical quadrature with a microwave field
acting as the meter \cite{hertzberg}.

\section{Experimental feasibility}
For the cavity electro-optic modulator reported by Ilchenko \textit{et
  al.}\ \cite{ilchenko},
\begin{align}
\omega_b &\approx 2\pi\times 9~\textrm{GHz},&
g &\approx 2\pi\times 20~\textrm{Hz},
\nonumber\\
\gamma_a &\approx 2\pi\times40~\textrm{MHz},
&
\gamma_b &\approx 2\pi\times 90~\textrm{MHz}.
\end{align}
If we assume a pump power $P$ of $2$~mW, $\lambda_0 \equiv 2\pi
c/\omega_a = 1550$~nm, and $\gamma_a^2/(16\delta^2)=0.5$, then
\begin{align}
|\alpha_-|^2&=\frac{\gamma_aP}
{\hbar\omega_a(\delta^2+\gamma_a^2/4)} \approx 1.7\times 10^8,
&
G &\approx 2\times 10^{-5}.
\end{align}
$G$ thus needs to be increased by a factor of $10^5$ for cooling to be
significant. Apart from increasing the pump power, one can also
improve the $g$ coefficient by reducing the size and capacitance of
the microwave resonator. If one can make use of the maximum
electro-optic coefficient $n^3 r\sim 300$ pm/V in lithium niobate
\cite{yariv}, make $d\sim 10$~$\mu$m instead of the 150~$\mu$m
reported in Ref.~\cite{ilchenko}, and assume $l/(c\tau)\sim 0.5$ and
$C\sim 1$~pF, the $g$ coefficient given by Eq.~(\ref{g}) can be as
high as $2\pi\times 5$~kHz, which would make $G \sim 0.3$. This $G$ is
already close to the upper limit $\gamma_a/\gamma_b\sim 0.4$ according
to Eq.~(\ref{G_limit}), so one should reduce $\gamma_b$ for more
significant cooling.  $\gamma_b$ can be significantly reduced if the
microwave resonator is unloaded or made of a better conductor.  The
quality factor of superconducting microwave resonators can be as high
as $2\times 10^6$ to $3\times 10^8$ \cite{microwave,day} and
$\gamma_b$ can then be reduced and $G$ be enhanced by a factor of
$10^4$ to $10^6$.  Hence, cooling and the other quantum effects
proposed in this paper should be experimentally observable in the near
future, as the technology for electro-optic integration continues to
improve.

\section{Conclusion}
In conclusion, I have outlined several interesting quantum effects
that can in principle be observed in electro-optic systems. Although
these effects are challenging to demonstrate experimentally with
current technology, they will enable new classical and quantum
information processing capabilities that should be useful for both
fundamental science and applied technology.

\section*{Acknowledgments}
Discussions with Tobias Kippenberg, Elanor Huntington, Matthew
Woolley, Carlton Caves, and Jeffrey Shapiro are gratefully
acknowledged. This work was supported in part by National Science
Foundation Grant Nos.~PHY-0903953 and 0653596 and Office of Naval
Research Grant No.~N00014-07-1-0304.

\end{document}